\documentclass[aip,reprint]{revtex4-1}

\usepackage{graphicx}
\usepackage{amsmath}
\usepackage{amssymb}

\usepackage{color}

\draft % marks overfull lines with a black rule on the right

\begin{document}

%\preprint{}

\title{Understanding migraine using dynamical network biomarkers}



\author{Markus A. Dahlem}
\email[]{Corresponding author:\\
Markus A. Dahlem, PhD\\
Tel: +49 (0)30 2093 99 185\\
Fax: +49 (0)30 2093 99188\\
Email: dahlem@physik.hu-berlin.de
}
%dahlem@physik.hu-berlin.de}
%\thanks{}
\affiliation{Department of Physics, AG NLD Cardiovascular Physics, Humboldt-Universit\"at zu Berlin, Robert-Koch-Platz 4, 10115 Berlin, Germany.}

\author{J\"urgen Kurths}
%\affiliation{Department of Physics, Humboldt-Universit\"at zu Berlin, Robert-Koch-Platz 4, 10115 Berlin, German}
\affiliation{Department of Physics, AG NLD Cardiovascular Physics, Humboldt-Universit\"at zu Berlin, Robert-Koch-Platz 4, 10115 Berlin, Germany.}
\affiliation{Potsdam Institute for Climate Impact Research, 14473 Potsdam, Germany.}
\affiliation{Institute for Complex Systems and Mathematical Biology, University of Aberdeen, Aberdeen AB24 3UE, United Kingdom.}

\author{Michel D. Ferrari}
\affiliation{Department of Neurology, Leiden University Medical Centre, Leiden, the Netherlands}
\author{Kazuyuki Aihara}

%\affiliation{Institute of Industrial Science, The University of Tokyo}
\affiliation{Collaborative Research Center for Innovative Mathematical Modelling, Institute of Industrial Science, University of Tokyo, Tokyo 153-8505, Japan.}

\author{Marten Scheffer}
\affiliation{Department of Aquatic Ecology \& Water Quality Management, Wageningen University, Wageningen, the Netherlands.}

\author{Arne May}
\affiliation{Center for Experimental Medicine,  Department of Systems Neuroscience, Universit\"atsklinikum Hamburg-Eppendorf, 20246 Hamburg, Germany.}


%\date{\today}

%Corresponding author:

%Markus A. Dahlem, PhD
%Department of Physics
%Humboldt-Universität zu Berlin
%Robert-Koch-Platz 4, 10115 Berlin, Germany
%Tel: +49 (0)30 2093 99 185
%Fax: +49 (0)30 2093 99188
%Email: dahlem@physik.hu-berlin.de

\begin{abstract}

\noindent {\bf Background}

\noindent Mathematical modeling approaches are becoming ever more established in clinical neuroscience. They provide insight that is key to understand complex interactions of network phenomena, in general, and interactions within the migraine generator network, in particular.

\noindent {\bf Purpose}

\noindent In this study, two recent modeling studies on migraine are set in the context of premonitory symptoms that are easy to confuse for trigger factors. This causality confusion is explained, if migraine attacks are initiated by a transition caused by a tipping point.

\noindent {\bf Conclusion}

\noindent We need to characterize the involved neuronal and autonomic subnetworks and their connections during all parts of the migraine cycle if we are ever to understand migraine. We predict that mathematical models have the potential to dismantle large and correlated fluctuations in such subnetworks as a dynamical network biomarker of migraine.
\end{abstract}


\maketitle

\section*{Bullet points}

\begin{itemize}
\item Article highlights the use of mathematical models in migraine research.

\item It explains causality confusion between triggers and premonitory symptoms by tipping points.

\item This explanation makes specific predictions of large scale correlated fluctuations that need to
be tested by noninvasive imaging.
\end{itemize}

\section*{Introduction}

Although migraine sufferers often are convinced that certain food, stress, bright light, neck pain, and other factors may trigger attacks, under controlled experimental conditions, there is very little if any evidence that these putative trigger factors can actually provoke attacks (1,2). Instead of being the trigger initiating an attack, craving for certain food, perceiving normal events as stressful or normal light intensities as too bright, and experiencing neck pain in the few hours to days prior to the clinical manifestation of an migraine attack more likely are early premonitory symptoms of an attack. Premonitory symptoms are actually expected as early--warning signs of an imminent transition and they are easy to confuse for trigger factors, if migraine attacks are initiated by a transition caused by a tipping point and therefore exhibiting universal behavior (3) (see Fig. 1). We addressed this question with mathematical migraine models in two recent articles (4, 5).

\begin{figure}
  \includegraphics[width=\columnwidth]{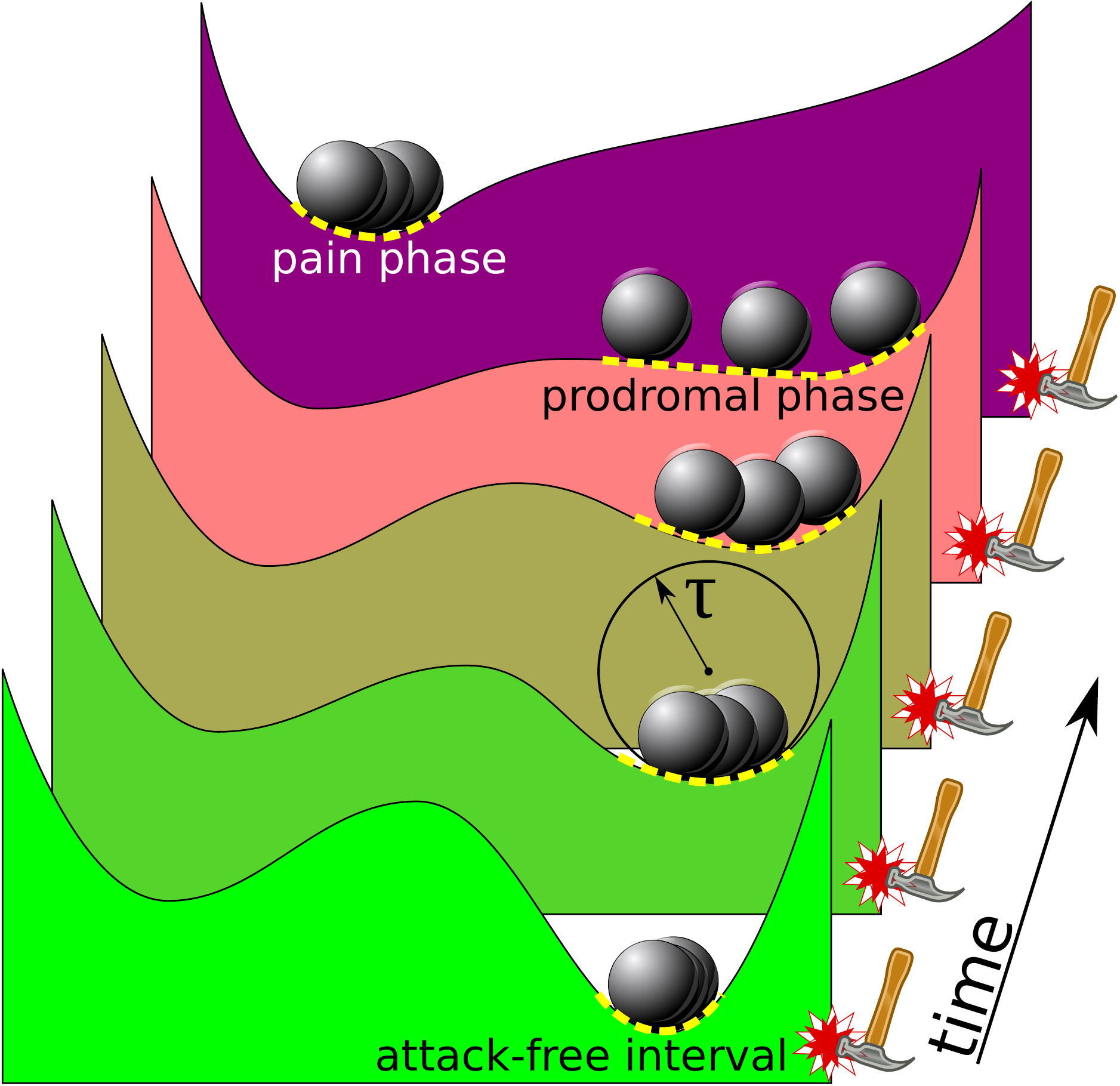}
\caption{Illustration of tipping point behavior: A changing landscape with potential wells and unique (violet landscape) or two (all other) stable attractors, representing the pain state (left) and attack--free state (right). The balls represent the current state and its variability (yellow dashed--line) under constant influence of noise as putative trigger factors (hammer). In this schematic illustration, a gradual increase of excitability over time lowers the potential well and the local landscape of the current state becomes shallower. The curvature of the well is inversely proportional to the system--immanent time scale ($\tau$) that determines the response to natural noise. The incipient loss of the threshold separating the attack--free minimum state from the attack state, that is, the prodromal phase, is indicated by both large amplitudes and critical slowing down (large $\tau$) of the fluctuations of the ball. These large amplitudes and critical slowing down of fluctuations can lead to confusing this for trigger factors, when in fact even purely internal noise (absence of hammer) will cause these eventually. Figure modified from Refs.\,(3,4).}
\end{figure}

\section*{Universal tipping point behavior explains premonitory symptoms}
The theoretical concept of tipping points, if transferred to migraine research, addresses the role of reduced resilience (4). As the brain comes to a tipping point, a small stimulus can trigger a slow cortical wave experienced as aura symptoms (6). The most important `cause' or `explanation'  of migraine as a chronic disease with episodic manifestations is therefore the dynamic change in excitability bringing the brain to this tipping point, rather than the small perturbation that finally tipped the balance leading to the attack as such. Although perturbations can be reduced, for example using exercise, stress relaxation techniques or cognitive behavioral therapy, they can not be excluded entirely near tipping points (Fig. 1). When the state of the brain gets again farther away from the tipping point, i.e., in a remissions phase of the migraine cycle, even very large perturbations---as recently published (1)---will not be able to trigger episodes. This refractoriness of hyperexcitability to external stimuli in the postictal and interictal phase can also be predicted as a general feature of chronic disorders with episodes caused by recurrently passing through a tipping point.

The neural correlates of early--warning signs caused by tipping points can be described as dynamical network biomarkers (DNB) (7). A DNB describes a certain behavior in a subnetwork of a complex disease, namely signals that announce the reduced resilience at an imminent tipping point by large and correlated amplitudes and critical slowing down of fluctuations in this subnetwork. In Fig. 1, this subnetwork is represented by the cross--section within a higher--dimensional landscape. Only in this particular cross--section the well is becoming shallower, while in all other directions perpendicular to this cross--section the well keeps its steep depth profile. DNB have been found for lung injury disease, liver cancer, and lymphoma cancer (7, 8). The theoretical concept and established methods of DNB can be transferred from such sudden deterioration diseases to chronic disorders with paroxysmal episodic manifestations like migraine (5).

\section*{Tipping points as a transdisciplinary concept}

Tipping points may be better known for the earth climate system (3). However, tipping points can be found in medicine, financial markets, traffic, power grid systems to which a large amount of renewable energy is introduced and that may fail therefore, ecosystems where wildlife populations may be threatened, and in the global climate system (9)---not too surprising, as all these are complex systems that exhibit nonlinear behavior and therefore are very likely to show tipping points.

Although mere coincidence, there are metaphors about migraine and the brain's climate, migraine being a thunderstorm or lightning in the head. It can be fruitful to see beyond such metaphors the consequences of tipping points and the related common structure of causal misinterpretation, both schematically illustrated in Fig. 1. Consider the statement of Kleinen et al. (10) referring to the North Atlantic currents : ``{\it It is becoming increasingly evident that there are critical thresholds in the Earth system, where the climate may change dramatically [...]. The exact positions of these thresholds are, however, still unclear and it might be doubted whether they can be determined with enough precision to give concrete information on the threat of crossing the threshold. Therefore, additional independent methods for assessing the closeness of the system to these thresholds are needed. These methods could contribute to an early warning system for assessing the danger of crossing a threshold and possibly provide the information necessary for controlling the system}'' (10). One arrives at a central question in migraine research that will profit from complex systems theory, when in this citation `Earth' is replaced with `brain' and `climate' with `neural dynamics' and when we start to use established concepts in climate research: How can we assess the proximity of a migraine threshold and, if the risk is large, control in this early stage the imminent migraine attack?

Climate change over decadal time scales probably involves changes in the ocean's conveyor belt, the thermohaline circulation, which was modeled in Ref.\ (10). Dahlem et al.\ (5) proposed to consider migraine pain caused by central sensitization in analogy as an overturning circulation in nerve traffic of the brain's migraine generator network (MGN) (11). What are early warning signals of this overturning circulation? Again, let us consider the climate system in analogy: It is easy to mistake cold winters for contradicting global warming, while in fact, severe winters like the ones of 2005-06 and 2009 do not conflict with the global warming picture, but rather supplement it as an integral part of the large and correlated amplitude fluctuations (12).

\section*{Neural correlate of premonitory symptoms in a subnetwork}
There are no simple answers to simple questions in nonlinear systems, in particular a causality interpretation is difficult. Reduced resilience and consequently large and slow fluctuations can explain the abovementioned situations where events that belong to the natural variability are mistaken near tipping points for triggers even if this is not intuitive for patients concerned. It was actually also suggested that migraine patients are driven or have the urge to exercise as a premonitory symptom (13). Unchallenged is that excessive yawning is a well know prodromal symptom in migraine, the same holds true for rapid mood changes, fatigue and craving for certain foods to name but a few. However, active coping, such as biofeedback (14) including contingent negative variation (15) but also behavioral treatments including relaxation trainings, stress--management training and cognitive--behavior therapy (16) clearly showed that the pre-transition state is in principle reversible, at least in some of the attacks. 

Where in the proposed subnetworks would such a behavioral therapy take its effects? Brainstem activation is thought to be specific for migraine attacks and specifically the dorsolateral pons has been repeatedly demonstrated by imaging data (17-19), while it also was suggested that this area alone cannot be the migraine generator (20). Given that the premonitory (21) and the subsequent attack symptoms (22) are characterized by interdependent networks which explains many of the facets of each event, the possible interplay between these networks has been coined the MGN (11).

If a DNB can be found in migraine it will identify the neural correlate for multiple early--warning signs as a common subnetwork of the MGN. In fact, there is a unitary hypothesis that identifies such a subnetwork---but only for multiple triggers causing migraine pain and strain (23). Therefore, if indeed triggers and symptoms are often mistaken at the incipient tipping point, this unitary hypothesis would suggest that large and correlated fluctuations in this subnetwork are crucial, that is, given the clinical picture, in the limbic system as well as the pre- and postganglionic parasympathetic neurons that control the sympathestic/parasympathestic balance. We predict therefore large and correlated fluctuations in this subnetwork as a DNB of migraine (5).

\section*{Conclusion}

To summarize, quantitative modeling approaches are becoming ever more established as a transdisciplinary research field. At the same time, the clinical research audience faces the difficult task, if not to penetrate mathematical concepts, at least to take away the message relevant for their own research. The particular message for clinical research is that our prediction must be tested: We need to characterize these neuronal, i.e., cortical, subcortical and autonomic subnetworks and their connections in the prodromal phase and the cortical slow wave  during the aura phase if we are ever to understand the true beginnings of an attack. The general message from complex systems theory is that migraine is an inherently dynamical disease (24) with a complex network generating interdependent events.

\section*{Acknowledgement}
Supported by the 7th Framework EU-project EuroHeadPain (\#602633) to AM and MF and the FIRST program from JSPS, initiated by CSTP to KA.

\section*{References}
{\footnotesize
\begin{enumerate}
\item Hougaard, A, Amin AFM, Amin F, Hauge AW, Ashina M, Olesen J. Provocation of
migraine with aura using natural trigger factors. Neurology 2013; 80: 428-431.

\item Dolgin E. Aura of mystery. Nat. Med. 2013; 19: 1083-1085.

\item Lenton TM, Held H, Kriegler E, Hall JW, Lucht W, Rahmstorf S, Schellnhuber HJ.
Tipping elements in the earth's climate system. Proc. Natl. Acad. Sci. USA 2008; 105:
1786-1793.

\item Scheffer M, van den Berg A, Ferrari MD. Migraine strikes as neuronal excitability
reaches a tipping point. PLOS ONE 2013; 8: e72514.

\item M. A. Dahlem, S. Rode, A. May, N. Fujiwara, Y. Hirata, K. Aihara, and J. Kurths,
Towards dynamical network biomarkers in neuromodulation of episodic migraine,
Transl. Neuroscie. 4, 282-294 (2013).

\item Charles AC, Baca SM. Cortical spreading depression and migraine. Nat. Rev. Neurol.
2013; Epub ahead of print.

\item Chen L, Liu R, Liu ZP, Li M, Aihara K. Detecting early-warning signals for sudden
deterioration of complex diseases by dynamical network biomarkers. Sci. Rep. 2012; 2:
342.

\item Liu R, Li M, Liu ZP, Wu J, Chen L, Aihara K. Identifying critical transitions and their
leading biomolecular networks in complex diseases. Sci. Rep. 2012; 2: 813.

\item Scheffer M, Bascompte J, Brock WA, Brovkin V, Carpenter RS, Dakos V, Held H, van
Nes EH, Rietkerk M, Sugihara G. Early-warning signals for critical transitions. Nature
2009; 461, 53-59.

\item Kleinen T, Held H, and Petschel-Held G, “The potential role of spectral properties in
detecting thresholds in the earth system: Application to the thermohaline circulation.
Ocean Dynamics 2003; 53: 53-63.

\item Dahlem MA. Migraine generator network and spreading depression dynamics as
neuromodulation targets in episodic migraine. Chaos 2013; 23: 046101.

\item Petoukhov V, Semenov VA. A link between reduced Barents-Kara sea ice and cold winter
extremes over northern continents. J. Geophys. Res. 2010; 115: D21111.

\item Goadsby PJ, Silberstein SD. Migraine triggers: harnessing the messages of clinical
practice. Neurology 2013; 80: 424-425.

\item Magis D, Schoenen J. Treatment of migraine: update on new therapies. Curr Opin
Neurol. 2011; 24:203-210.

\item Kropp P, Gerber WD. Contingent negative variation--findings and perspectives in
migraine. Cephalalgia. 1993; 13:33-36.

\item Penzien DB, Andrasik F, Freidenberg BM, Houle TT, Lake AE 3rd, Lipchik GL, Holroyd
KA, Lipton RB, McCrory DC, Nash JM, Nicholson RA, Powers SW, Rains JC, Wittrock
DA. American Headache Society Behavioral Clinical Trials Workgroup.; Guidelines for
trials of behavioral treatments for recurrent headache, first edition: American Headache
Society Behavioral Clinical Trials Workgroup. ; Headache. 2005; 45 Suppl 2:S110-32.

\item Weiller C, May A, Limmroth V, J\"uptner M, Kaube H, Schayck R v, Coenen HH, Diener
HC. Brain stem activation in spontaneous human migraine attacks; Nature Medicine,
1995; 1: 658- 660.

\item Stankewitz A, Aderjan D, Eippert F, May A. Trigeminal nociceptive transmission in
migraineurs predicts migraine attacks. J. Neurosci., 2011; 31:1937-1943.

\item Goadsby PJ, Fields HL. On the functional anatomy of migraine.; Ann Neurol. 1998
43:272.

\item Borsook D, Burstein R. The enigma of the dorsolateral pons as a migraine generator.
Cephalalgia 2012; 32: 803-812.

\item Maniyar FH, Sprenger T, Monteith T, Schankin C, Goadsby PJ. Brain activations in the
premonitory phase of nitroglycerin-triggered migraine attacks. Brain. 2014 137:232-241.

\item Stankewitz A, Aderjan D, Eippert F, May A. Trigeminal nociceptive transmission in
migraineurs predicts migraine attacks. J. Neurosci 2011; 31:1937-1943.

\item Burstein R, Jakubowski M. Unitary hypothesis for multiple triggers of the pain and strain
of migraine. J. Comp. Neurology 2005; 493: 9-14.

\item Mackey MC, Milton JG. Dynamical diseases. Ann. N. Y. Acad. Sci. 1987; 504: 16-32.

\end{enumerate}

}
%\section*{References}
%\bibliography{ref}

\end{document}